\documentclass[twocolumn,twocolappendix]{aastex63}

\usepackage[T1]{fontenc}
\usepackage{ae,aecompl}

\usepackage{natbib}
\usepackage{multirow}
\usepackage{diagbox}

\usepackage{footnote}
\usepackage[flushleft]{threeparttable} 
\usepackage{rotating}

\usepackage{amsmath}
\usepackage{upgreek}
\usepackage{xfrac}  

\newcommand\mc[1]{\multicolumn{1}{c}{#1}}

\shorttitle{The dependence of RLF of QSOs on redshift}
\shortauthors{Rusinek-Abarca \& Sikora}

\begin{document}

\title{The dependence of the fraction of radio luminous quasars on redshift and
its theoretical implications}

\correspondingauthor{Katarzyna Rusinek-Abarca}
\email{krusinek@camk.edu.pl}

\author[0000-0002-6424-6558]{Katarzyna Rusinek-Abarca}
\affiliation{Nicolaus Copernicus Astronomical Center, Polish Academy of Sciences, Bartycka 18, 00-716 Warsaw, Poland}

\author[0000-0003-1667-7334]{Marek Sikora}
\affiliation{Nicolaus Copernicus Astronomical Center, Polish Academy of Sciences, Bartycka 18, 00-716 Warsaw, Poland}


\begin{abstract}
While radio emission in quasars can be contributed to by a variety of processes (involving star forming regions, accretion disk coronas and winds, and jets), the powering of the radio loudest quasars must involve very strong jets, presumably launched by the Blandford-Znajek mechanism incorporating the magnetically arrested disk (MAD) scenario. We focus on the latter and investigate the dependence of their fraction on redshift. We also examine the dependence of the radio-loud fraction (RLF) on BH mass ($M_{\rm BH}$) and Eddington ratio ($\lambda_{\rm Edd}$) while excluding the redshift bias by narrowing its range. In both these investigations we remove the bias associated with:
(1) the diversity of source selection by constructing two well-defined, homogeneous samples of quasars (first within $0.7 \leq z < 1.9$, second within $0.5 \leq z < 0.7$);
(2) a strong drop in the RLF of quasars at smaller BH masses by choosing those with BH masses larger than $10^{8.5} M_{\odot}$.
We confirm some previous results showing the increase in the fraction of radio-loud quasars with cosmic time and that this trend can be even steeper if we account for the bias introduced by the dependence of the RLF on BH mass whereas the bias introduced by the dependence of the RLF on Eddington ratio is shown to be negligible. Assuming that quasar activities are triggered by galaxy mergers we argue that such an increase can result from the slower drop with cosmic time of mixed mergers than of wet mergers.
\end{abstract}


\section{Introduction}
\label{sec_1_INTRODUCTION}

As recent studies suggest, the quasi-steady production of strong, relativistic jets in certain quasars over $10^7-10^8$ years \citep[e.g.][]{Bird2008}, with some jet powers approaching their accretion power \citep{vanVelzenFalcke2013,Rusinek2017,Inoue2017}, may require the formation of magnetically arrested disks (MADs). Their formation takes place if the net accumulated magnetic flux exceeds the maximal amount which can be confined on the black hole by the accretion flow \citep{Bisnovatyi-KoganRuzmaikin1976,Narayan2003,Igumenshchev2008,McKinney2012}. However, the proposed scenarios which build up MADs and explain why they occur in only a fraction of quasars remain in the realm of speculation. Their uncertainties are mainly associated with our poor knowledge about the magnetic properties (intensities, topology, and diffusivity) of accretion flows and their dependence on initial conditions which trigger quasar activity \citep[see e.g.][]{Begelman1995,Beckwith2008}. The most widely accepted scenarios include: 
\begin{itemize}
\item[(A)] the possibility of building-up the MAD by the advection of the poloidal magnetic fields by accretion flows \citep{Lubow1994,Spruit2005,Bisnovatyi-KoganLovelace2007,RothsteinLovelace2008,Beckwith2009,GuiletOgilvie2012,GuiletOgilvie2013,Cao2016,CaoLai2019};
\item[(B)] the formation of the MAD in situ, e.g.,~via the ''cosmic battery'' \citep{ContopoulosKazanas1998,KoutsantoniouContopoulos2014}; 
\item[(C)] the survival of the MAD, formed during the thick, hot, very low accretion rate phase, during the transition of to the quasar phase \citep{Sikora2013,SikoraBegelman2013}.
\end{itemize}

Observationally, one can try to confront the above scenarios by studying: the differences between multi-band spectra of radio-loud (RL) and radio-quiet (RQ) quasars; the differences between properties of their hosts and environments; and the dependence of the radio-loud fraction (RLF) of quasars on redshift.

The most striking result of comparing the radiative properties of quasars is that the spectral energy distributions (SEDs) of RL and RQ quasars are very similar \citep[e.g.][]{Elvis1994,Richards2006,deVries2006,Shang2011,Shankar2016}. Statistically significant differences are noticed only in the X-ray bands. In particular, RLQs are found to be on average X-ray louder and with X-ray spectra harder than RQQs. Whilst the X-ray hardness may result from the dependence of the X-ray spectra slopes on the Eddington ratio and the lower average Eddington ratio of RL quasars than of RQ quasars, the difference in the X-ray loudness requires something else. Assuming that the production of X-rays is powered by reconnection of magnetic fields in accretion disk coronas \citep{Beloborodov2017,Sironi2020}, the efficiency of X-ray production in the vicinity of the BH is likely to be differentiated by the presence/absence of the MAD. Obviously, structures of the innermost portions of accretion flows also depend on values of BH spin. However, noting that many radio-quiet AGNs have very large spins \citep[][and refs.~therein]{Reynolds2019}, the precise value of the spin is not expected to be directly correlated with the X-ray production efficiency. Hence, the similarity of SEDs between RL and RQ quasars for similar Eddington ratios, BH masses, and spins precludes the existence of MAD from being dependent on the current accretion parameters.
    
Contrary to the radiative properties, severe differences between RL and RQ quasars are established regarding their hosts and environment, the main ones being:
\begin{itemize}
\item[(1)] in opposition to RQ quasars which are found to be hosted by both, elliptical (E) and spiral (Sp) galaxies, the vast majority of RL quasars are hosted by giant ellipticals \citep[gE; e.g.][]{Floyd2010,Tadhunter2016,Rusinek2020};
\item[(2)] the RL quasars are located in denser environments and have on average much more massive dark matter halos than RQ quasars \citep[e.g.][]{Mandelbaum2009,Shen2009,Donoso2010,Wylezalek2013,Retana-Montenegro2017}.
\end{itemize}

Adopting the premise that the triggering of the high accretion events represented by the quasar phenomenon is associated with galactic mergers \citep[e.g.][]{Shen2009,Bessiere2012,Treister2012}, one might expect that almost all radio-loud quasars and some fraction of radio-quiet quasars are triggered by mergers of massive elliptical galaxies with disk galaxies, while the remaining RQ quasars formed from mergers of two disk galaxies. Within this scheme the question remains what decides the quasar radio loudness following the merger of a massive elliptical galaxy with a disk galaxy. As was argued by \citet{Rusinek2020}, this can be explained by involving the ''cosmic-battery'' scenario of MAD formation, provided the merger leads to a BH system which is either corotating, or counterrotating while satisfying the stability criterion derived by \citet{King2005}. However such a duality can be achieved also within the 'C' scenario, because not all MADs formed during the geometrically thick, hot, very low accretion rate phase of MAD can survive the transition of the object to the quasar phase \citep[see Figure 5 in][]{Rusinek2017}.

The possible dependency of MAD formation on the type of galaxy mergers can be verified by comparing the cosmic history of different types of mergers with the dependence of RLF of quasars on redshift. As semi-analytical models indicate \citep{Khochfar2003} and observations support \citep{Lin2008}, the fraction of E-Sp mergers decreases with increasing redshift, while the fraction of Sp-Sp mergers increases with redshift. Then, noting that the dominant fraction of RQ quasars is presumably triggered by the Sp-Sp mergers, while the vast majority of RL quasars -- by E-Sp mergers, one might expect to have RLF decreasing with redshift. Unfortunately the results of studies on such a dependence performed in the past often conflict with each other \citep[see e.g.][]{Stern2000,Jiang2007,Singal2011,Singal2013,Kratzer2015}. The reasons for this are biases associated with the radio and optical flux limits which work differently for differently selected samples and the choice of the demarcation value of the radio loudness parameter used to divide quasars into radio-loud and radio-quiet catagories.

In order to minimize the effects of the aforementioned biases, we adopt the following strategy:
\begin{itemize}
\item[--] we study the dependence of the RLF on redshift for quasars selected from the sample analyzed by \citet{Gurkan2019}, distinguishing three subsamples, altogether covering a redshift range of $0.7 \leq z < 1.9$ (see Section~\ref{sec_2_RLFvsz});
\item[--] we investigate the dependence of the RLF on BH mass and Eddington ratio for a well-defined, homogeneous, and  redshift-narrowed ($z \sim 0.6$, thus excluding the dependence of the RLF on the redshift) sample for quasars chosen from the catalog of \citet[][see Section~\ref{sec_3_RLFvsMBHandEddratio}]{Shen2011};
\item[--] we define the RLF as the ratio of RL to the total number of sources where RL objects are classified as those in which the radio emission is dominated by the presence of strong jets -- most likely produced within the MAD scenario;
\item[--] we choose in the above samples sources with BH masses larger than $10^{8.5} M_{\odot}$ in order to avoid biases associated with a large drop in the fraction of RLQ at smaller BH masses.
\end{itemize}

The possible theoretical interpretations of the  anticorrelation of the RLF with redshift, the dependence of the RLF on BH mass, $M_{\rm BH}$, and lack thereof with regards to Eddington ratio, $\lambda_{\rm Edd}$, that we found are discussed in Section~\ref{sec_4_DISCUSSION} \citep[such a dependence is observed for QSOs with $\lambda_{\rm Edd} < 0.01$ and is caused by the decrease of the radiative efficiency of the accretion flows at such low Eddington ratios,][]{Sikora2007}. The main results of our work are summarized in Section~\ref{sec_5_SUMMARY}.

Throughout the paper we assume a $\Lambda$CDM cosmology with $H_0 = 70 \, {\rm km} \, {\rm s}^{-1} \, {\rm Mpc}^{-1}$, $\Omega_m=0.3$, and $\Omega_{\Lambda}=0.70$.

\section{The dependence of radio-loud fraction of quasars on redshift}
\label{sec_2_RLFvsz}

With a focus on performing a study of the radio-loud fraction of quasars on redshift, we decide to use sources from \citet{Gurkan2019} who collected 49 925 optically selected quasars for detailed radio analysis. The authors made use of the sensitive and high-resolution low-frequency radio data, for which the extended radio structures of an AGN (lobes, plumes, etc.) dominate the radio emission with the Doppler boosting being minimized. Together with the wide range of redshifts, BH masses, and Eddington ratios that the objects analyzed by \citet{Gurkan2019} cover (see Table 3 and Figure 8 therein), their sample enabled us to construct a well-controlled, homogeneous population, a thorough description of which is given below.

Objects analyzed by \citet{Gurkan2019} are chosen from the SDSS-IV DR14 quasar catalog (DR14Q). Compiled from the extended Baryon Oscillation Spectroscopic Survey (eBOSS) of SDSS IV by \citet{Paris2018} and based on works of \citet{Myers2015} and \citet{Blanton2017}, DR14Q contains 526 356 quasars and provides redshifts, photometry in five bands, and information about broad absorption line quasars as well as, when available, multi-wavelength matching with large-area surveys. 

\citet{Gurkan2019} focused on quasars from the Hobby-Eberly Telescope Dark Energy Experiment \citep[HETDEX,][]{Hill2008} Spring field and Herschel-Astrophysical Terahertz Large Area Survey/North Galactic Pole \citep[H-ATLAS/NGP,][]{Hardcastle2016} region for which data obtained by Low Frequency Array \citep[LOFAR,][]{vanHaarlem2013} as a part of the LOFAR Two-metre Sky Survey \citep[LoTSS,][]{Shimwell2017} are accessible. The advantage of using LoTSS relates to its high sensitivity ($\sim 100\,\upmu$Jy) which combined with a resolution of 6\,arcsec and the low frequency it has been conducted at, $120-168$\,MHz, allows for the detection of the extended emission often missed at higher frequencies, at which most of the commonly used radio surveys have been carried out (including the Faint Images of the Radio Sky at Twenty-cm catalog which we used, see Section~\ref{subsubsec_3.1.3_RLFvsMBHandEddratio_sample_radiodata}). Thereby, the sample analyzed by \citet{Gurkan2019} is the largest set of optically selected quasars detected at $144$\,MHz to date.

Based on the availability and the character of the LOFAR data, \citet{Gurkan2019} divide their quasars into those with and without radio detections, distinguishing compact and extended objects among the former. Unfortunately we cannot use their classification for our analysis, as this information is not provided elsewhere than within the figures presented in their work. From their Figure 4 which shows the radio loudness distribution, with the radio loudness parameter being the ratio of monochromatic luminosities at $144$\,MHz, $L_{\nu_{\rm 144}}$, to SDSS $i$-band, $L_{\nu_i}$, in a logarithmic scale, we can see that while all radio-loud objects (corresponding to those with $L_{\nu_{\rm 144}}/L_{\nu_i} > 10^3$, in accordance to the division introduced by \citeauthor{Kellermann1989}~\citeyear{Kellermann1989}, see Appendix~\ref{appendix_A_conversionR}) have direct radio measurements, a significant fraction of radio-intermediate (RI, with values of $10^2 < L_{\nu_{\rm 144}}/L_{\nu_i} < 10^3$) ones lack those. Furthermore, one can note that sources with extended radio emission, albeit representing about $5\%$ of all detected objects (865 out of 16 259), occupy quite a wide range of radio loudness values, from 1 up to $10^4$.

\begin{table*}[ht!]
\begin{center}
\caption{Comparison of the total numbers of sources and RL objects only, $\rm RLF$ values, median redshifts, BH masses, and Eddington ratios for three samples selected from \citet{Gurkan2019}.}
\label{tbl_1_RLF_redshift}
\hspace{-1.5cm}
\begin{tabular}{ccccccc} \hline
Sample & $N_{\rm total}$ & $N_{\rm RL}$ & $\rm RLF$ & $\tilde{z}$ & $\log \tilde{M}_{\rm BH} [M_{\odot}]$ & $\log \tilde{\lambda}_{\rm Edd}$\\ \hline \hline
$0.7 \leq z < 1.1$  & 2 172 & 79 & \ $3.64\%$ & $0.93$ & $8.79$ & $-1.20$ \\ 
$1.1 \leq z < 1.5$  & 3 210 & 85 & \ $2.65\%$ & $1.31$ & $8.89$ & $-0.96$ \\ 
$1.5 \leq z < 1.9$  & 3 795 & 73 & \ $1.92\%$ & $1.69$ & $8.95$ & $-0.88$ \\ \hline
\end{tabular}
\end{center}
\end{table*}

While selecting quasars from the sample of \citet{Gurkan2019} we decided on a redshift limit of $0.7 \leq z < 1.9$ and a BH mass cutoff of $M_{\rm BH} \geq 10^{8.5} M_{\odot}$. The first criterion provided us with a sufficiently large sample size (at lower redshifts the number of sources drops significantly) over a wide enough range of redshifts to perform our study and, more importantly, was chosen so that it coincides with uniform calculations of BH masses and bolometric luminosities, $L_{\rm bol}$, which, within the given redshift range, are estimated from the Mg\,II line and monochromatic luminosity at $3000$\,\AA~(values of $M_{\rm BH}$ and $\lambda_{\rm Edd}$ were taken from \citeauthor{Shen2011}~\citeyear{Shen2011}, and \citeauthor{Kozlowski2017}~\citeyear{Kozlowski2017})\footnote{Even though we rejected sources with different estimations of $M_{\rm BH}$ and $L_{\rm bol}$ than mentioned here, we note that the calculations made by \citet{Shen2011} and \citet{Kozlowski2017} may vary slightly from each other. These discrepancies, however, do not affect our further analysis as they are smaller than the reported uncertainties of $M_{\rm BH}$ and $L_{\rm bol}$, being $\sim 0.06$ vs.~$\sim 0.4$ dex and $\sim 0.01$ vs.~$\sim 0.1$ dex, respectively (see \citeauthor{Kozlowski2017} \citeyear{Kozlowski2017} for more details).}. The second constraint, as already explained, gives possibly the largest ratio of radio-loud to total objects within the sample. Such a sample contains 9 177 quasars..

We further split our sample into three groups on the basis of their redshift, $0.7 \leq z < 1.1$, $1.1 \leq z < 1.5$, and $1.5 \leq z < 1.9$, amounting to 2 172, 3 210, and 3 795 objects, respectively. The radio-loud fractions computed for each of these populations were found to be $3.64\%$, $2.65\%$, and $1.92\%$, accordingly, clearly decreasing with increasing redshift. The opposite happens with the medians of BH mass and Eddington ratio, which both increase with redshift\footnote{We checked that the trend was preserved while considering the effect of the measurement uncertainties of $M_{\rm BH}$ and $L_{\rm bol}$ which are typically about $0.4$ and $0.1$ dex, respectively. The uncertainty of $\lambda_{\rm Edd}$, being $\sim 0.41$ dex, was obtained as described in Section~\ref{subsubsec_3.1.2_RLFvsMBHandEddratio_sample_MBHLbolEddratio}.}. The size of each sample, number of RL objects, median redshifts, BH masses, and Eddington ratios are provided in Table~\ref{tbl_1_RLF_redshift}.

\section{The dependence of radio-loud fraction of quasars on BH mass and Eddington ratio}
\label{sec_3_RLFvsMBHandEddratio}

As the dependence of the radio-loud fraction of quasars on redshift can be affected not only by the distinct cosmic history of triggers of radio-loud and radio-quiet quasars but also by the dependence of the RLF on the BH mass and the Eddington ratio and the correlation of the median values of $M_{\rm BH}$ and $\lambda_{\rm Edd}$ with redshift (see Table~\ref{tbl_1_RLF_redshift}), we decide to investigate such a~dependence in more detail. However, to conduct such a~study, the analyzed sample of sources has to be chosen from a very narrow range of redshift while being quite numerous, especially regarding the ratio of radio-loud to total objects with as few sources lacking radio data among those with the highest values of radio loudness as possible. For these reasons the previously analyzed subsamples from \citet{Gurkan2019} are not sufficient as their RLF is rather small and it drops even further when subject to narrower redshift ranges. Thus a new sample of quasars located closer has been chosen -- with a narrow range of redshift of $0.5 \leq z < 0.7$.

\subsection{The Sample}
\label{subsec_3.1_RLFvsMBHandEddratio_sample}

\subsubsection{Source Selection}
\label{subsubsec_3.1.1_RLFvsMBHandEddratio_sample_selection}

The objects collected in our studied sample of quasars at $z \sim 0.6$ are initially taken from the catalog of quasar properties from Sloan Digital Sky Survey Data Release 7 (SDSS DR7) compiled by \citet{Shen2011} from works of \citet{Schneider2010} and \citet{Abazajian2009}. Besides careful analysis of the continuum and emission line measurements around the H$\upalpha$, H$\upbeta$, Mg\,II, and C\,IV regions, the authors provide information about radio properties, flags indicating quasars with broad absorption lines or disk emitters, and black hole masses. We distinguish four main steps which lead to our final sample (hereinafter: $0.5-0.7$ QSOs): narrowing the redshift range; checking whether BH masses and bolometric luminosities (and with those -- Eddington ratios) are estimated in the same manner for all the objects; limiting BH masses; and choosing sources in the footprint of Faint Images of the Radio Sky at Twenty-cm \citep[FIRST,][]{Becker1995} catalog.

The catalog from \citet{Shen2011} gathers 105 783 quasars within a wide range of redshifts (from the very nearby Universe up to $z=5.5$). We selected from them quasars enclosed within a narrow range of redshifts of $0.5 \leq z < 0.7$\footnote{The redshifts used in our work are taken, just as in \citet{Shen2011}, from the SDSS DR7 quasar catalog \citep{Schneider2010}. These values do not differ significantly from the more accurate data provided by \citet{HewettWild2010} and consequently do not affect our final sample size and the obtained results.} leaving us with 7 306 objects. Such a~restraint provides the following advantages: (1) it minimizes the redshift bias from the dependence of the RLF on BH mass and Eddington ratio; and (2) the chosen redshift range achieves a suitable trade-off between minimizing the biases stemming from the optical and radio flux limits while avoiding a~significant reduction in the sample size.

The previous constraint already indicates homogeneous data that should be used for the estimation of $M_{\rm BH}$ and $\lambda_{\rm Edd}$. We verify this information (Section~\ref{subsubsec_3.1.2_RLFvsMBHandEddratio_sample_MBHLbolEddratio}) and exclude quasars with measurements obtained differently which reduces the sample size by 97 sources.

In the next step, like in case of studied by us sample from \citet{Gurkan2019}, we limited the sample for $M_{\rm BH} \geq 10^{8.5} M_{\odot}$ which further restricts our sample to 3 734 objects.

Finally, as the subject of our research is to examine the radio-loud fraction of quasars, we reject all the objects which are located outside of the FIRST footprint. This step sets the number of our final sample to 3 511 quasars. The reason for choosing FIRST as our main source of radio data and the way it was assigned to each quasar is explained in Section \ref{subsubsec_3.1.3_RLFvsMBHandEddratio_sample_radiodata}. 

Table \ref{tbl_2_sample_constraints} summarizes how each of the above-mentioned steps contributed to the final size of the $0.5-0.7$ QSOs sample.

\begin{table}[t]
\begin{center}
\caption{Subsequent steps involved in constructing our $0.5-0.7$ QSOs sample.}
\label{tbl_2_sample_constraints}
\begin{tabular}{cl} \hline
Number of objects & Constraint \\ \hline
105 783 & Sample from \citet{Shen2011} \\ 
\ \ \ 7 306 & Redshift cutoff of $0.5 \leq z < 0.7$ \\ 
\ \ \ 7 209 & Uniform calculation of $M_{\rm BH}$ and $\lambda_{\rm Edd}$ \\ 
\ \ \ 3 734 & Limit of $M_{\rm BH} \ge 10^{8.5} M_{\odot}$ \\
\ \ \ 3 511 & Objects within the FIRST footprint \\ \hline
\end{tabular}
\end{center}
\end{table}

\subsubsection{Black Hole Mass, Bolometric Luminosity, and Eddington Ratio}
\label{subsubsec_3.1.2_RLFvsMBHandEddratio_sample_MBHLbolEddratio}

As stated in \citet{Shen2011}, the fiducial virial black hole mass for quasars at $z < 0.7$ is estimated from the broad H$\upbeta$ line following the calibration provided by \citet{VestergaardPeterson2006} while the bolometric luminosity for those quasars is computed from the monochromatic luminosity at $5100$\,\AA \ using the spectral fits and bolometric corrections from \citet{Richards2006}. We examined whether the sources we collected follow these two conditions and rejected those whose $M_{\rm BH}$ and $L_{\rm bol}$ are obtained from other measurements assuring that all the calculations for our quasars are alike.

The BH masses of objects in our sample have their values within the range of $8.5 \leq \log M_{\rm BH} \, \rm{[M_{\odot}]} \leq 10.3$ with median of $\log \tilde{M}_{\rm BH}\, \rm{[M_{\odot}]} = 8.81$. Uncertainties for each BH mass estimate are typically  $\sigma_{\log M_{\rm BH}} \sim 0.4$\,dex.

Because the bolometric luminosities provided by \citet{Shen2009} are overestimated by including the contribution from infrared radiation (see \citealt{Marconi2004} and footnote 19 in \citealt{Shen2011}) we reduce their bolometric luminosities by one third accounting for this IR emission.  

The values of $L_{\rm bol}$ for objects in our sample are within a range of $44.10 \leq \log L_{\rm bol} \, \rm{[erg\,s^{-1}]} \leq 47.09$ with a median of $\log \tilde{L}_{\rm bol} \, \rm{[erg\,s^{-1}]} = 45.54$. The typical uncertainty of the bolometric luminosity measurement is $\sigma_{\log L_{\rm bol}} \sim 0.1$\,dex.

At last we compute the Eddington ratio, $\lambda_{\rm Edd} \equiv L_{\rm bol}/L_{\rm Edd}$, finding its values enclosed within the range of $-3.30 \leq \log \lambda_{\rm Edd} \leq -0.07$ with a median of $\log \tilde{\lambda}_{\rm Edd} = -1.39$. The typical uncertainty of the Eddington ratio is estimated as $\sigma_{\log \lambda_{\rm Edd}} = \sqrt{\sigma_{\log M_{\rm BH}}^2 + \sigma_{\log L_{\rm bol}}^2}$ which gives $\sim 0.41$ when using the typical uncertainties mentioned above.

\subsubsection{Radio Data}
\label{subsubsec_3.1.3_RLFvsMBHandEddratio_sample_radiodata}

Even though the radio properties are included in \citet{Shen2011}, their procedure of assigning radio data is not scrupulous enough with regard to the objective of our study, being particularly biased against extensive double structures. Due to this we carry out the matching procedure following the strategy adopted by \citet{Rusinek2020}.

The radio catalog we used is FIRST, which was designed to overlap with the area of SDSS. This 1.4\,GHz sky survey is characterized by its high resolution (5.4\,arcsec) and its sensitivity down to 1\,mJy\footnote{Despite the fact that FIRST has about 10 times worse sensitivity than LoTSS, it is deep enough to proceed with our study of quasars located at $z \sim 0.6$, for which radio structures above $\sim 33$\,kpc are resolved.}.

At first we conducted a search within a matching radius of 1' of the optical position. For sources where only one radio association was uncovered, if the radio location was found to be within 5'' of the optical position, the match was confirmed  and the object was designated as \textit{'compact'}. Objects with multiple associations were examined manually through visual inspection of radio maps with sizes of 0.45\,deg $\times$ 0.45\,deg extracted from FIRST\footnote{http://sundog.stsci.edu/}. If, from visual examination, it was determined that the object was not compact, i.e.~multiple radio matches were found to be associated with the object, then the source was classified as \textit{'extended'}.  Among them, we distinguish those in which a pair of lobes was determined, referring to them as \textit{'lobed'} (regardless of the detection of the radio core). For quasars in which much more extended, i.e.~beyond 1', radio morphologies were noticed, the search for radio matches was gradually increased by 1' until the whole structure was identified. The radio flux from each component was then summed up for every extended source.

Sources with non detections were assigned an upper limit of 1\,mJy corresponding to the detection limit of FIRST. 

Only 16\% of $0.5-0.7$ QSOs QSOs are radio detected (546 out of 3 511) and 31\% of those reveal extended radio morphology (171 out of 546). The radio lobed objects constitute 80\% of the radio extended ones (136 out of 171)\footnote{We note that our assignment and categorization of the FIRST radio data, despite being cautiuos, is more likely incomplete rather than contaminated with false associations as we reject any radio matches with unconvincing linkage to the quasar which is exactly the opposite to the approach of \citet{Shen2011} who assigned all the nearby (i.e.~located within their 30'' matching radius) associations as authentic.}.

\subsection{Radio Loudness Distribution} 
\label{subsec_3.2_RLFvsMBHandEddratio_radioloudness}

\begin{figure}[t!]
\begin{center}
\includegraphics[width=0.48\textwidth]{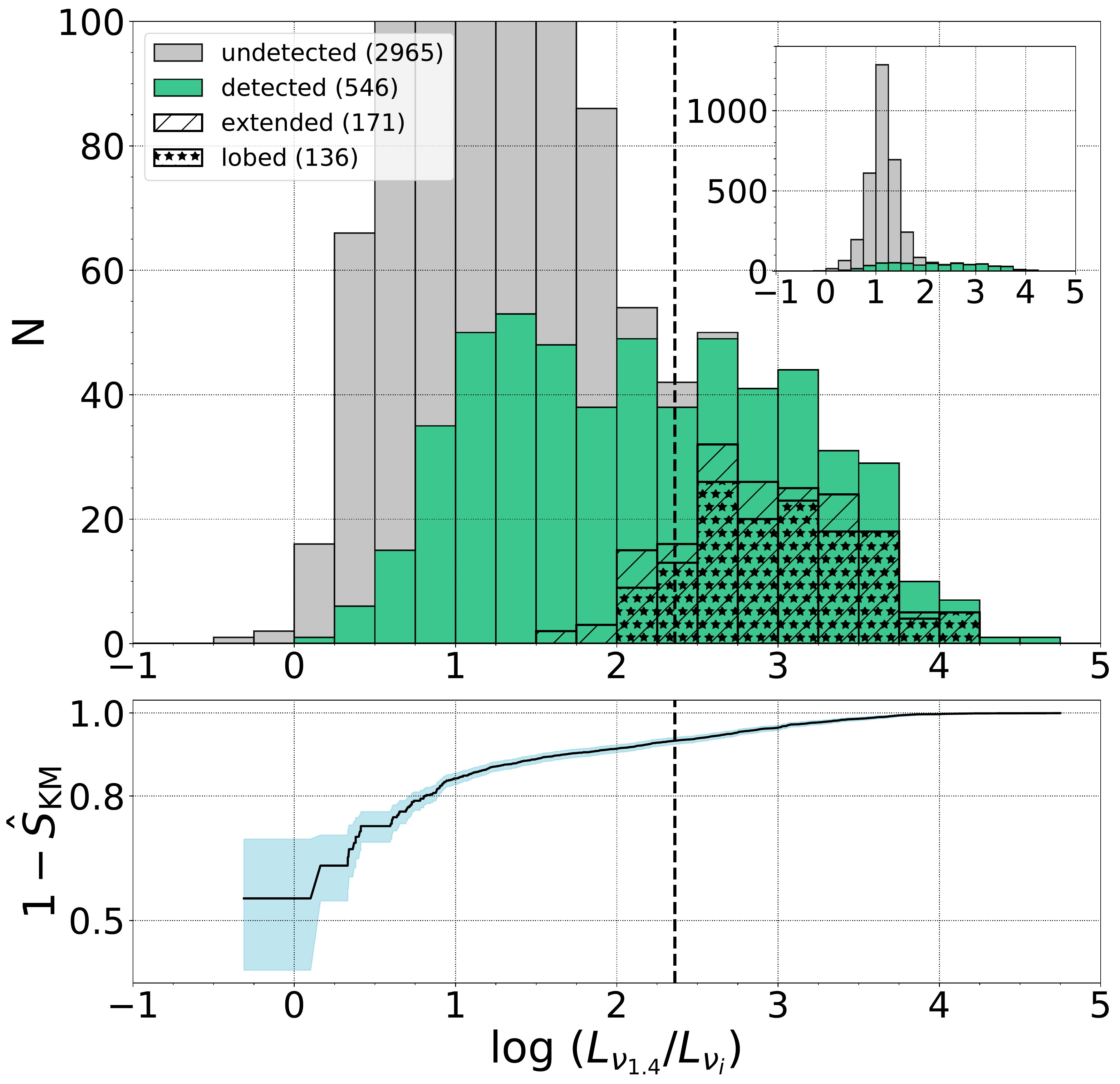}
\caption{The top panel presents the radio loudness distribution for our $0.5-0.7$ QSOs sample. Sources with and without radio detections are shown separately (in green and grey, accordingly). Within radio detected objects we distinguish those with radio extended morphologies (hatched) and those with morphologies that exhibit two clearly separated lobes (starred). The black dashed line marks the value of $\log\,(L_{\nu_{1.4}}/L_{\nu_i}) = 2.36$ separating objects into RL and RI\&RQ, following the definition of radio loudness by \citet{Kellermann1989}. The characteristics of these radio morphologies are closely described in Section \ref{subsubsec_3.1.3_RLFvsMBHandEddratio_sample_radiodata}. The bottom panel depicts the estimated cumulative distribution function (black solid line) taking into acount left-censored data points uing the Kaplan-Meier estimator for the survical function \citep{KaplanMeier1958}. The light blue area ilustrates the $95\%$ confidence interval for the estimator.}
\label{fig_1_hist_logR}
\end{center}
\end{figure}

In order to separate objects with powerful jets dominating the radio emission from those in which this emission is contributed to mainly from other processes, we use the radio loudness parameter and define it as the ratio of monochromatic luminosities at $1.4$\,GHz, $L_{\nu_{1.4}}$, to, similarly to \citet{Gurkan2019}, SDSS $i$-band, $L_{\nu_i}$. In Appendix~\ref{appendix_A_conversionR} we present the general formula for the conversion of the radio loudness parameter between various frequencies referring to the definition introduced by \citet{Kellermann1989}.

The radio loudness distribution of our $0.5-0.7$~QSOs is presented in top panel of the Figure~\ref{fig_1_hist_logR}. Its values spread over four orders of magnitude, $-0.31 \leq \log\,(L_{\nu_{1.4}}/L_{\nu_i}) \leq 4.75$, with its peak at $\log\,(L_{\nu_{1.4}}/L_{\nu_i}) \sim 1.2$ which, as one can see, is strongly dominated by the radio undetected sources which constitute the vast majority of the whole sample. On the other hand, the radio detected sources are much more evenly distributed over the whole range of radio loudness with the median being about ten times higher than for the radio undetected ones ($144.41$ and $13.87$, accordingly).

\begin{table}
\begin{center}
\caption{Radio morphologies and radio classes of all $0.5-0.7$~QSOs. A detailed description is given in Sections~\ref{subsubsec_3.1.3_RLFvsMBHandEddratio_sample_radiodata} and \ref{subsec_3.2_RLFvsMBHandEddratio_radioloudness}.}
\label{tbl_3_radio_classes_morphology}
\begin{tabular}{lrcccc} \hline
\multicolumn{2}{c}{\multirow{2}{*}{Radio Morphology}} & \multicolumn{3}{c}{Radio Class} & \multirow{2}{*}{Total}  \\ \cline{3-5}
\multicolumn{2}{c}{} & RL & RI & RQ &  \\ \hline \hline
\multirow{3}{*}{Detected} & Extended & 145 & 26 &  & 171 \\  
   &  (Lobed) & (122) & (14) &  & (136) \\ 
& Compact &  88 & 160 & 127 &  375 \\ \hline
 Undetected  &  & 2 & 509 & 2454 & 2965  \\ \hline \hline
\multicolumn{2}{c}{Total}  & 235 & 695 & 2581 & 3511  \\ \hline
\end{tabular}
\begin{flushleft}
\textbf{Note.} Lobed sources are a subgroup of extended ones. Their exact counts, which are shown in brackets, are not added to the total since they are already included in the numbers for extended sources.
\end{flushleft}
\end{center}
\end{table}

\begin{figure*}
\begin{center}
\includegraphics[width=0.7\textwidth]{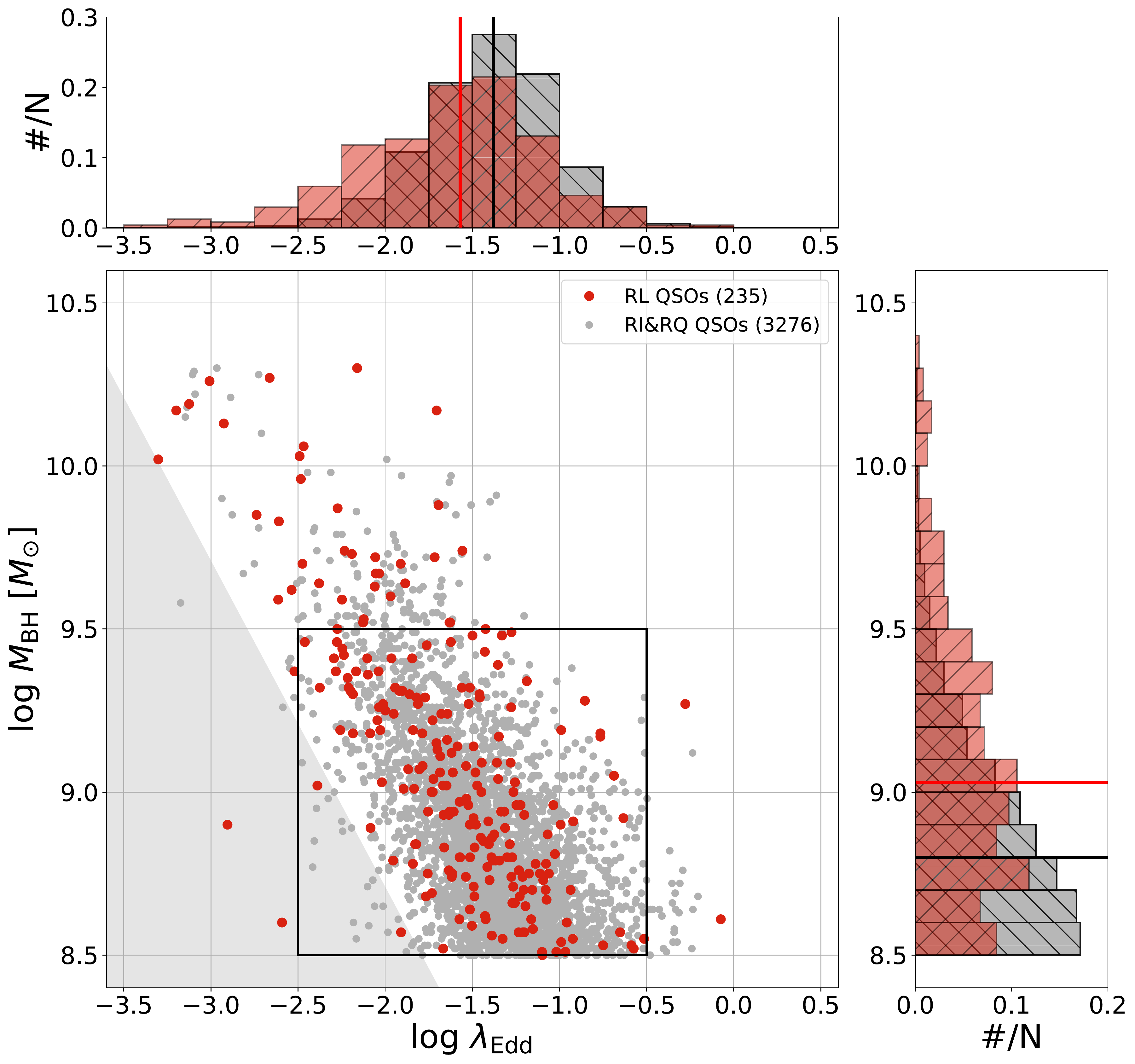}
\caption{Distribution of all the $0.5-0.7$~QSOs in $\log \lambda_{\rm Edd}$ vs.~$\log M_{\rm BH}$ plane. Radio-loud and radio-intermediate with radio-quiet quasars are shown separately (in red and grey, accordingly). The shaded area on the lower-left corner marks the region of sources fainter than those observed by SDSS, corresponding to $L_{\rm bol, min} \simeq \sfrac{2}{3} \times 10^{45} \rm{erg\,s^{-1}}$ (at $z \sim 0.6$ and $M_i \leq -22$), and for which the estimation of Eddington ratio is biased. The best representation of RL to RI\&RQ objects overlaps with the limits of the inlaid box which are further adopted for Figure~\ref{fig_3_logEddratio_logMBH_RLF}. The median values of $\log \lambda_{\rm Edd}$ and $\log M_{\rm BH}$ presented on the relevant histograms are: $\log \lambda_{\rm Edd, RL} = -1.57$ and $\log \lambda_{\rm Edd, RI\&RQ} = -1.38$; $\log M_{\rm BH, RL} = 9.03$ and $\log M_{\rm BH, RI\&RQ} = 8.80$ (all marked in red for RL and black for RI\&RQ).}
\label{fig_2_logEddratio_logMBH}
\end{center}
\end{figure*}

The bottom panel of the Figure~\ref{fig_1_hist_logR} shows the Kaplan-Meier estimator for the cumulative distribution function (CDF), $\hat{F}$, which takes into account the upper limits used for the radio-undetected sources \citep{KaplanMeier1958}. The estimator was computed using the python package \texttt{lifelines} \citep{davidson2019lifelines} to esimate the survival function $\hat{S}_\mathrm{KM} = 1 - \hat{F}$. The upper panel might suggest that there are almost no sources with $\log\,(L_{\nu_{1.4}}/L_{\nu_i)} <0$ and maybe half the sources with $\log\,(L_{\nu_{1.4}}/L_{\nu_i}) \lesssim1$, however, the estimated CDF shows that likely half or more of the sources should have $\log\,(L_{\nu_{1.4}}/L_{\nu_i}) <0$ and over $80\%$ should have $\log\,(L_{\nu_{1.4}}/L_{\nu_i})<1$. The blue shaded area represents the $95\%$ confidence interval for the estimator. For higher values of the radio loudness the estimated CDF has very low uncertainty because almost all of these sources correspond to real measurements.

Dividing objects into radio-loud, radio-intermediate, and radio-quiet (based on the shape of the radio loudness distribution and the demarcation introduced by \citeauthor{Kellermann1989}~\citeyear{Kellermann1989} being RL for ${\mathcal R}_{\rm K} > 100$, RI for $10 < {\mathcal R}_{\rm K} < 100$, RQ when ${\mathcal R}_{\rm K} < 10$; see Appendix~\ref{appendix_A_conversionR} for recalculation between $\mathcal{R}_{\rm K}$ and $L_{\nu_{1.4}}/L_{\nu_i}$), we find 235, 695, and 2 581 sources belonging to the given class, respectively. Such a categorization corresponds well with the presence or lack of detectable radio emission and its character. Extended, and among them lobed, sources are almost exclusively radio-loud spanning the two highest decades of radio loudness. This group contains less than $1\%$ of radio undetected objects{\footnote{Among the 235 RL sources, only two objects do not have assigned radio data. Those are SDSS J003330.69+004251.4, with $\log M_{\rm BH} = 8.6$, $\log \lambda_{\rm Edd} = -2.59$ and SDSS J125303.76+402749.9 with $\log M_{\rm BH} = 8.9$, $\log \lambda_{\rm Edd} = -2.91$.}} which is exactly the opposite for radio-quiet sources being dominated by objects lacking radio detections. Finally, quasars classified as radio-intermediate have compact, extended, and lobed morphologies but also no radio data assigned. Specific numbers of the above-mentioned relations are listed in Table \ref{tbl_3_radio_classes_morphology}.

\subsection{Radio-Loud Fraction}
\label{subsubsec_3.3_RLFvsMBHandEddratio_RLF}

Knowing the radio loudness distribution of $0.5-0.7$~QSOs and splitting them into RL and RI\,\&\,RQ groups, we place them on the $\lambda_{\rm Edd}$ -- $M_{\rm BH}$ plane shown in Figure~\ref{fig_2_logEddratio_logMBH}. Both groups occupy similar ranges of both parameters with RL objects being more massive and having lower accretion rates than RI\,\&\,RQ ones. The limit of $M_{\rm BH} \ge 10^{8.5} M_{\odot}$ especially affects the radio-quieter sources as their BH mass distribution is smoothly increasing towards lower values of $M_{\rm BH}$.

The lack of quasars in the lower-left corner of the diagram is a consequence of the SDSS DR7 optical flux limit in the $i$-band, while the deficiency of quasars in the upper-right corner is associated with the downsizing effect \citep[e.g.][]{Fanidakis2012,Hirschmann2014}. Most of the objects are located within an area of $-2.5 \leq \log \lambda_{\rm Edd} \leq -0.5$ and $8.5 \leq \log M_{\rm BH} \leq 9.5$ providing the best representation of RL to RI\,\&\,RQ objects upon which the radio-loud fraction is analyzed (and within which all the radio-loud sources are radio-detected).

\begin{figure}[t]
\begin{center}
\includegraphics[width=0.48\textwidth]{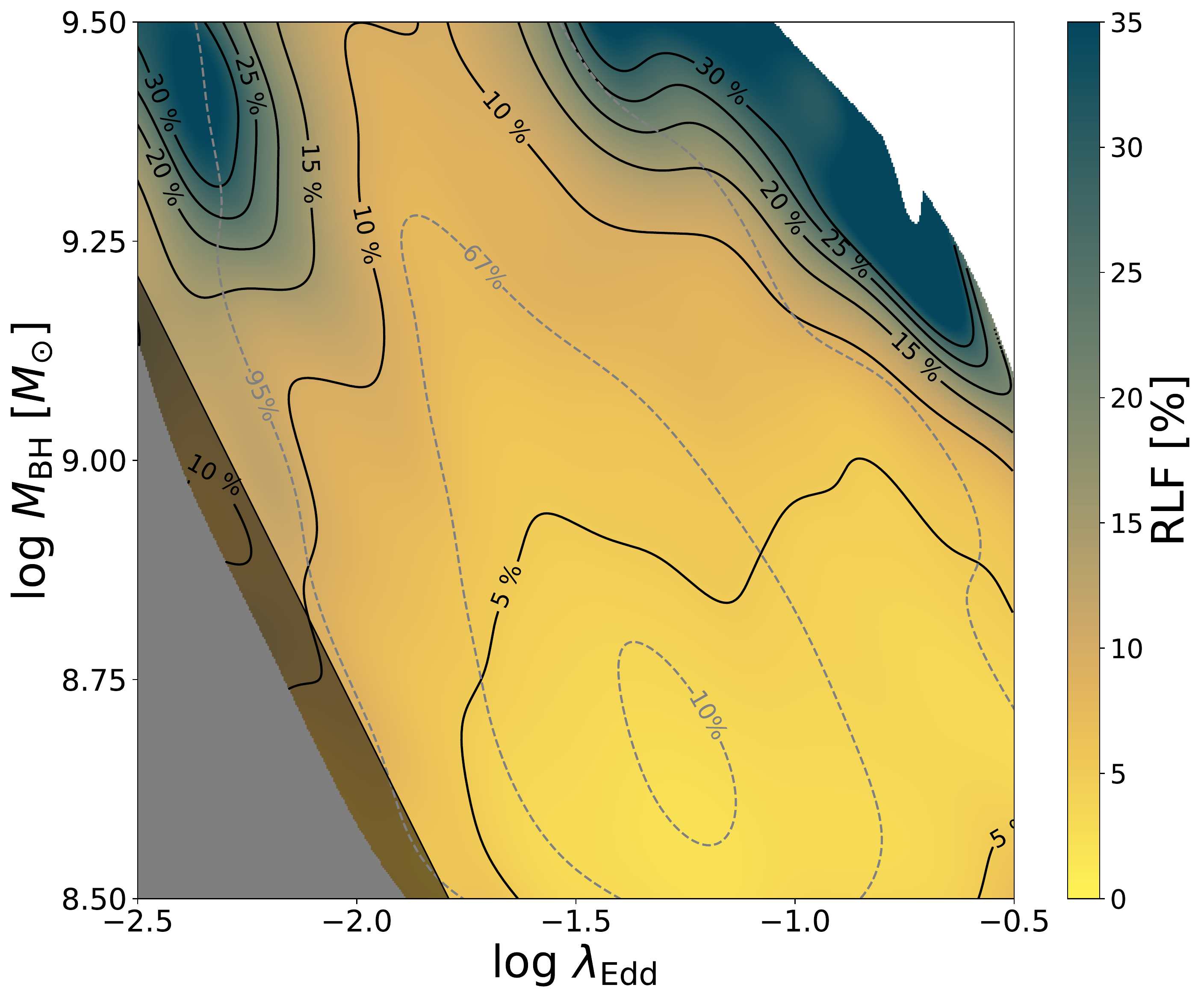}
\caption{The radio-loud fraction as a function of $\log \lambda_{\rm Edd}$ and $\log M_{\rm BH}$ computed from a gaussian kernel density estimate for the number density of radio loud and radio quiet sources. The shaded area on the left corner demonstrates the region of sources with a biased estimation of Eddington ratio (just as in Figure~\ref{fig_2_logEddratio_logMBH}), while the white corners, bottom left and top right, are masked due to the low number density of RL as well as RI\&RQ objects (below 10). The RLF is represented with the color density map and the explicit percentages are shown on the black solid contours. The grey dashed contours correspond to isodensity contours enclosing the indicated percentages of the total sample.}
\label{fig_3_logEddratio_logMBH_RLF}
\end{center}
\end{figure}

Since the probability of production of powerful jets by quasars can depend not only on the redshift, but also on the BH mass and on the specific accretion rate (the latter being traced for a given radiative efficiency of the accretion flow by the Eddington ratio), the RLF -- defined to be the ratio of the RL quasars to all quasars -- is expected to depend on the distribution of the BH masses and Eddington ratios in the sample. Indeed, as we can see in Figure~\ref{fig_2_logEddratio_logMBH}, the number of sources strongly varies in $M_{\rm BH}$ and $\lambda_{\rm Edd}$. Thus we study the dependence of the RLF of quasars on the BH mass and Eddington ratio within a limited range of both parameters as presented in Figure~\ref{fig_3_logEddratio_logMBH_RLF}.

Figure~\ref{fig_3_logEddratio_logMBH_RLF} demonstrates the increase of RLF with BH mass and its near lack of a dependence on the Eddington ratio. Some variations among the fixed values of $M_{\rm BH}$ and $\lambda_{\rm Edd}$ are apparent but this should not be surprising after considering the number of sources in specific regions. And as we can see, despite these variations, the correlation of RLF with BH mass is clearly visible at all Eddington ratio values, while no correlation between RLF and Eddington ratio can be inferred at any of the BH mass values. Such a trend still persists while considering the non negligible effect of the measurement uncertainties on the distributions of $M_{\rm BH}$ and $\lambda_{\rm Edd}$.

\section{Discussion} 
\label{sec_4_DISCUSSION}

Quasars can be divided into those with radio production dominated by jets and those with dominant radio contribution coming from processes not associated with the presence of jets. A particular subclass of the first population are quasars associated with classical FR\,II radio sources \citep{FanaroffRiley1974}. The distribution of their radio loudness defined by \citet{Kellermann1989}, ${\mathcal R}_{\rm K} = L_{\nu_5}/L_{\nu_{\rm B}}$ (see Appendix~\ref{appendix_A_conversionR}), spans the range $10^2-10^4$, with a peak at $\sim 10^3$, while the radio of the second population is presumably dominated by star forming regions \citep[SFRs, e.g.][]{Condon1992,Kimball2011,Panessa2019,Koziel-Wierzbowska2021}, with their radio loudness distribution enclosed within the range of $0.1 < {\mathcal R}_{\rm K} < 10$ (see Figure~\ref{fig_1_hist_logR}). With such representations, they cover well separated ranges of ${\mathcal R}_{\rm K}$ and form a clear bimodal distribution, with the first population called RLQs and the second population called RQQs. However not all ${\mathcal R}_{\rm K} > 100$ quasars are FR\,II, others have more complex radio morphology or are unresolved. Part of them fill up the gap in ${\mathcal R}_{\rm K}$-histograms between quasars with ${\mathcal R}_{\rm K} > 10$ and quasars with ${\mathcal R}_{\rm K} < 100$ and represent the so-called radio-intermediate quasars. Since they are too radio-loud to be explained by processes taking place in SFRs, they are often considered to have radio emission powered by jets, but with most of the jet energy dissipated within galaxies like those represented by FR\,I quasars \citep{Blundell2001,Heywood2007}. However ${\mathcal R}_{\rm K}$ with values up to tens are achievable also by radio emission powered by accretion disk winds \citep{Blundell2007,Zakamska2014,Nims2015,Zakamska2016,White2017MNRAS,Hwang2018,Morabito2019,Rosario2020,Vayner2021}.  Hence, with the possible presence of such quasars in the radio-intermediate band ($10 < {\mathcal R}_{\rm K} < 100$), the commonly used definition in the literature of the RLF as the ratio of objects with ${\mathcal R}_{\rm K} > 10$ to their total number may lead to a significant overestimation of the fraction of quasars which have radio emission dominated by jet activity. Noting that such an overestimation can depend on the redshift and that within the RI band there are also objects with upper limits of ${\mathcal R}_{\rm K}$ (see Table~\ref{tbl_3_radio_classes_morphology}), we focused our studies on the RLF of quasars with ${\mathcal R}_{\rm K} > 100$, where they all are 'radio-categorized' (in the sense of not including quasars with only upper limits on the radio flux) and are expected to be fully represented by quasars with strong jets.

The negative correlation of radio-loud fraction with redshift we identified in Section~\ref{sec_2_RLFvsz} can be examined towards lower values of redshift. For this purpose we consider the well-established sample of $0.5-0.7$~QSOs introduced in Section~\ref{sec_3_RLFvsMBHandEddratio} (with RLF being $6.69\%$) and an additional set of AGNs located nearby compiled by \citet{Rusinek2020}. The advantage of using the latter arises not only from the great fraction of radio detected sources studied therein (257 out of 314 objects, $82\%$) but also from the similar approach of building their sample, i.e.~the estimation of BH masses and bolometric luminosities (hence -- Eddington ratios), which was performed in the same manner for all the sources, the limitation of $M_{\rm BH} \geq 10^{8.5} M_{\odot}$, the usage of radio data at 1.4\,GHz (collected from i.a.~FIRST), and the division of the sample into RL, RI, and RQ objects, as well as those with and without radio detections. The main differences are the source selection, which in the case of \citeauthor{Rusinek2020} was done from the \textit{Swift}/Burst Alert Telescope (BAT) AGN Spectroscopic Survey catalog \citep[BASS,][]{Ricci2017} while our $0.5-0.7$~QSOs and the samples selected from \citet{Gurkan2019} are based on optical data from SDSS,  and the calculation of radio loudness as \citeauthor{Rusinek2020} uses infrared instead of optical data to trace the accretion power choosing the W3 band, $\lambda_{\rm W3} = 12\,\upmu \rm m$\footnote{As it was demonstrated by \citet{Ichikawa2019}, mid-IR radiation in nearby luminous AGNs is dominated by dust heated by radiation from the accretion disk.}, from Wide-field Infrared Survey Explorer (WISE, \citeauthor{Cutri2013} \citeyear{Cutri2013}. The reasoning behind this choice is provided in Appendix A.2 in \citeauthor{Rusinek2020} \citeyear{Rusinek2020}; see also \citeauthor{Gupta2020} \citeyear{Gupta2020}). Such a well-defined sample was studied regarding its radio and optical morphologies, radio loudness distribution, and jet production efficiencies by \citet{Rusinek2020} and analyzed earlier by \citet{Gupta2018,Gupta2020} in the context of the similarities and differences of the spectral energy distributions between RL and RQ AGNs of both Type 1 and Type 2.

\begin{figure}[t]
\begin{center}
\includegraphics[width=0.48\textwidth]{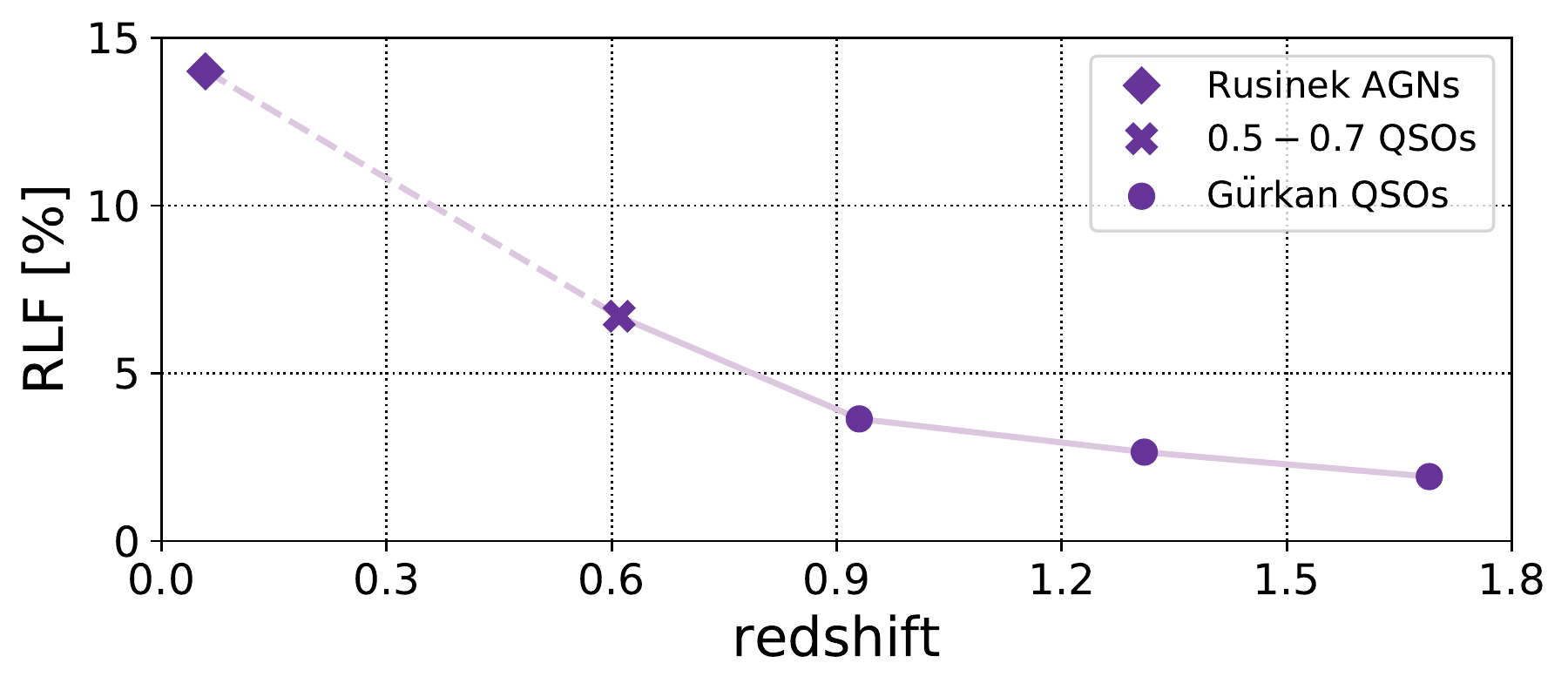}
\caption{The radio-loud fraction as a function of redshift. The five various samples correspond to: the \textit{Swift}/BAT AGN sample taken from \citet{Rusinek2020} represented by a diamond;  the set of $0.5-0.7$ QSOs discussed in Section~\ref{sec_3_RLFvsMBHandEddratio} being marked as a cross; and three populations of quasars selected from \citet{Gurkan2019} which are shown as circles. The specific values of RLF and median redshifts are: $14.00\%$, $6.69\%$, $3.64\%$, $2.65\%$, $1.92\%$; and $0.06$, $0.61$, $0.93$, $1.31$, $1.69$, respectively. These values should be treated with caution, as an outline of the dependence we found between those two parameters (RLF and $z$) due to some differences between the samples. Especially given that the sources in \citeauthor{Rusinek2020} were selected based on X-ray instead of optical data.}
\label{fig_4_RLF_redshift_5samples}
\end{center}
\end{figure}

The relation between the RLF and redshift with additional two sets of sources is shown in Figure~\ref{fig_4_RLF_redshift_5samples}. As one can clearly see, the previously found trend is maintained down to the lowest redshifts covered by \textit{Swift}/BAT AGNs for which the RLF is found to be $14\%$. The real dependence of RLF on the redshift is expected to be even steeper as it may be affected by the correlation of RLF with BH masses (as shown in Table~\ref{tbl_1_RLF_redshift} and studied thoroughly in  Section~\ref{sec_3_RLFvsMBHandEddratio}). In addition to that, the radio frequency at which the given sample was observed would also contribute to the steeper dependence of RLF on redshift, i.e.~some of the radio-quieter sources in our $0.5-0.7$ QSOs and \textit{Swift}/BAT AGNs taken from \citet{Rusinek2020} could reveal their more extended radio structures with their radio luminosity increasing while being observed at lower frequencies, whilst the opposite would be observed for radio-loud objects from the sample of \citet{Gurkan2019} seen at higher radio frequencies. On the other hand, the RLF in the AGN sample taken from \citet{Rusinek2020} can be slightly overestimated due to the fact that this sample was selected based on X-ray observations and that RLQs are found to be on average X-ray louder than RQQs \citep[see][]{Gupta2018,Gupta2020}.

Even though the specific values of RLF presented in Figure~\ref{fig_4_RLF_redshift_5samples} should be treated with caution, the negative correlation of the radio-loud fraction with redshift we found suggests differential evolution of the space density of triggers of jetted and non-jetted quasars, with the former dropping with cosmic time slower than the latter. In both cases the quasar triggers are likely to be associated with galaxy mergers \citep[e.g.][]{Lin2008,Shen2009,Bessiere2012,OLeary2021}. As studies of optical morphologies of nearby RL and RQ objects indicate, those leading to triggering of radio-loud quasars presumably involve mergers of giant ellipticals with gas-rich disk galaxies, while most of radio-quiet quasars are born following mergers of two disk galaxies \citep[e.g.][]{Rusinek2020}. Such a connection between quasar types and merger types is also supported by studies of the redshift evolution of mergers which show that the ratio of E-Sp merger events to Sp-Sp merger events is increasing with cosmic time \citep[see][and refs.~therein]{Lin2008}. However there is also a non-negligible fraction of RQ quasars which, like RL quasars, are hosted by gE, and which, like RL ones, are presumably triggered by mergers of gE with disk galaxies.


\subsection{On the origin of the diversity of the jet production efficiency in RL quasars}
\label{subsec_4.1_DISCUSSION_diversityjetproduction}

Using radio lobe calorimetry to calculate jet powers in FR\,II quasars \citep[e.g.][]{Willott1999,ShabalaGodfrey2013} and optical or infrared data to calculate bolometric luminosities of the accretion flow \citep[see][and refs.~therein]{Richards2006} allows for the estimation of the jet production efficiency, $\eta_j = P_j / \dot{M} c^2 = \epsilon_d \, P_j / L_{\rm bol}$, where $P_j$ is the jet power and $\epsilon_d = L_{\rm bol} / \dot{M} c^2$ is the radiative efficiency of the accretion disk with $\dot{M}$ being the accretion rate. According to \citet{vanVelzenFalcke2013}, \citet{Inoue2017}, and \citet{Rusinek2017} this efficiency is spread over about two decades, with the median on the order of $0.01 (\epsilon_d/0.1$). Such efficiencies are achievable in the magnetically arrested disk (MAD) models, according to which $P_j \sim a^2 \dot M c^2 (H/R)$, where $-1 < a < 1$ is the dimensionless BH spin and $(H/R)$ is the geometrical thickness of the accretion flow \citep[see review by][]{Davis2020}.


\subsection{Can MADs represent temporary states during the quasar lifetime?}
\label{subsec_4.2_DISCUSSION_QSOlifetime}

Any intermittency of production of jets powering FR\,II radio structures should result in having some FR\,IIs without radio cores which are known to represent Doppler-boosted parcsec-scale portions of the jet \citep[e.g.][]{Blandford1979,Marin2016}. Since radio cores are observed in all FR\,II quasars \citep[e.g.][]{vanGorkom2015}, such an intermittency is rather excluded. However, this does not exclude the possibility of strong modulation of the jet production, which can be responsible for knotty structures of some large scale jets \citep[see e.g.][]{Godfrey2012}. Such modulations can be explained in terms of the MAD model by noting that according to this model the power of a jet scales with the accretion rate and this, in turn, can be modulated by the viscous instabilities in the accretion disk \citep{Janiuk2011}.

Claims about lacking an intermittency in the production of powerful jets might be contradictory to having too many luminous Gigahertz-Peaked Spectrum (GPS) sources if they are considered to have powerful jets at young ages \citep{Reynolds1997}. However, multi-frequency monitorings suggest that such an excess is presumably a consequence of treating many blazars as young, unresolved double radio sources \citep{Mingaliev2012,Sotnikova2019}.


\subsection{Why should MADs only be available in quasars activated by mergers involving massive ellipticals and why not in all of them?}
\label{subsec_4.3_DISCUSSION_mergers}

An answer for this double question can be provided in terms of the scenario marked in the Section~\ref{sec_1_INTRODUCTION} by 'C'. According to this scenario, the MAD is formed during the quasi-spherical, low accretion rate phase. The presence of such a phase accompanied by jet activity is indicated by observations of radio activity in the nuclei of massive elliptical galaxies \citep{Sabater2019}. Depending on such parameters as the accretion rate and the size of the formed MAD, some MADs can survive transition to the quasar phase, while others cannot (see \citealt{SikoraBegelman2013} and Figure~5 in \citealt{Rusinek2017}).

Another possibility may involve scenario 'B', according to which the merger may lead to the formation of co- and counterrotating accretion flows relative to the BH spin \citep[see][and refs.~therein]{Garofalo2020}. As argued by \citet{Rusinek2020}, in the latter case the operation of the Poynting-Robertson process can be efficient enough to lead to the formation of a MAD. However it is not clear whether such counterrotating BH-disk systems can be stable \citep{King2005}.

In both the above scenarios the probability of getting MAD correlates with the BH mass, which is consistent with the relation between RLF and BH mass found by \citet{Kratzer2015} and confirmed by us. Having RQ quasars hosted not only by disk galaxies but also by elliptical galaxies weakens the differentiation of the dependence of the RLF of RQ and RL quasars on redshift, but not significantly, if the majority of RQ quasars are hosted by disk galaxies.

\section{Summary}
\label{sec_5_SUMMARY}

We studied the dependence of the radio-loud fraction (RLF) of quasars with BH masses larger than $10^{8.5} M_{\odot}$ on:
\begin{itemize}
\item[--] redshift, using three well-defined samples within the range of $0.7 \leq z < 1.9$ selected from the SDSS DR14 catalog of quasars matched with the LoTSS radio catalog;
\item[--] black hole masses and accretion rates, using a highly controlled, and redshift-narrowed (around $z \sim 0.6$) sample selected from the SDSS DR7 catalog of quasars matched with the FIRST radio catalog,
\end{itemize}
while treating as radio-loud objects those with radio loudness parameter defined by \citet{Kellermann1989} to be larger than $100$. In the first step we found that the RLF decreases with increasing redshift (see Table~\ref{tbl_1_RLF_redshift} and Figure~\ref{fig_4_RLF_redshift_5samples} with the addition of two more samples in the latter).

A detailed investigation of the relations between the RLF and BH masses and Eddington ratios, as well as a more general study of the radio properties, was conducted with the sample of quasars with redshifts enclosed within the range of $0.5 \leq z < 0.7$. This allowed us to verify whether and how the dependence of the RLF on $M_{\rm BH}$ and $\lambda_{\rm Edd}$ can bias the dependence of the RLF on redshift that we found. Our results show that the dependence of the RLF on redshift is expected to be biased only by the dependence of the RLF on BH mass. Since the average BH masses in the \citeauthor{Gurkan2019} sample increase with redshift (see Table~\ref{tbl_1_RLF_redshift}), such a bias weakens the dependence of the RLF on redshift, i.e.~without that bias the drop of the RLF with redshift would be even steeper than presented in the Figure~\ref{fig_4_RLF_redshift_5samples}.

Assuming that quasar activities are triggered by mergers of galaxies and that, contrary to radio-quiet quasars, the vast majority of RL quasars is hosted by giant ellipticals, we showed that the decreasing fraction of RL quasars with redshift can simply result from the theoretically predicted and observationally confirmed slower drop with cosmic time of mergers of giant ellipticals with spiral galaxies than of mergers of two spiral galaxies. Finally, we speculate about possible scenarios of formation of magnetically arrested disks (MADs), which could explain the connections indicated by observations between types of mergers and the production of powerful jets.

Our studies should be treated as a starting point for further more extensive analysis which could be carried out using e.g.~the upcoming LoTSS DR2 catalog (Schimwell et al., in prep.), which covers a sky area of $5720$ deg$^2$, i.e.~over 13 times bigger than a sky area of LoTSS DR1 used by \citet{Gurkan2019}.

\acknowledgments

We thank David Abarca for useful discussions, editorial and linguistic
assistance, and technical advice. We thank the anonymous referee for his attentive comments which significantly contributed to improving the
quality of the publication. The research leading to these results has received funding from the Polish National Science Centre grant 2015/18/A/ST9/00746.

\textit{Facilities:} Funding for the Sloan Digital Sky Survey IV has been provided by the Alfred P. Sloan Foundation, the U.S. Department of Energy Office of Science, and the Participating Institutions. 

SDSS-IV acknowledges support and resources from the Center for High Performance Computing  at the University of Utah. The SDSS website is www.sdss.org.

SDSS-IV is managed by the Astrophysical Research Consortium for the Participating Institutions of the SDSS Collaboration including the Brazilian Participation Group, the Carnegie Institution for Science, Carnegie Mellon University, Center for Astrophysics | Harvard \& Smithsonian, the Chilean Participation Group, the French Participation Group, Instituto de Astrof\'isica de Canarias, The Johns Hopkins University, Kavli Institute for the Physics and Mathematics of the Universe (IPMU) / University of Tokyo, the Korean Participation Group, Lawrence Berkeley National Laboratory, Leibniz Institut f\"ur Astrophysik Potsdam (AIP), Max-Planck-Institut f\"ur Astronomie (MPIA Heidelberg), Max-Planck-Institut f\"ur Astrophysik (MPA Garching), Max-Planck-Institut f\"ur Extraterrestrische Physik (MPE), National Astronomical Observatories of China, New Mexico State University, New York University, University of Notre Dame, Observat\'ario Nacional / MCTI, The Ohio State University, Pennsylvania State University, Shanghai Astronomical Observatory, United Kingdom Participation Group, Universidad Nacional Aut\'onoma de M\'exico, University of Arizona, University of Colorado Boulder, University of Oxford, University of Portsmouth, University of Utah, University of Virginia, University of Washington, University of Wisconsin, Vanderbilt University, and Yale University.

\textit{Software:} The \texttt{python} package \texttt{lifelines} was used to compute the Kaplan-Meier estimator \citep{davidson2019lifelines}.

\appendix
\section{Conversion of radio loudness parameter}
\label{appendix_A_conversionR}

Among the various methods used to classify quasars into RL or RQ, to this day the most common one is still the approach described by \citet{Kellermann1989}, defining the radio loudness parameter as the ratio of radio-to-optical monochromatic luminosity, at $5$\,GHz and in B-band, being
\begin{equation}
\label{eq_A1_RKdefinition}
{\cal R}_{\rm K} = L_{\nu_5} / L_{\nu_{\rm B}} =  (F_{\nu_5} / F_{\nu_{\rm B}}) (1+z)^{\alpha_{\rm r}-\alpha_{\rm o}},
\end{equation}
where $F_{\nu_5}$ and $F_{\nu_{\rm B}}$ are the flux densities at $\nu_5 \equiv 5$\,GHz and $\nu_{\rm B} \simeq 6.8 \times 10^{14}$\,Hz with $\alpha_{\rm r}$ and $\alpha_{\rm o}$ being the spectral radio and optical indeces, given as $\alpha = -\frac {d \ln F_{\nu}}{d \ln \nu}$, respectively. Then the radio loudness defined at any radio frequency, $\nu_{\rm r}$, and optical frequency, $\nu_{\rm o}$ (alternatively, at infrared, $\nu_{\rm IR}$, or X-ray frequency, $\nu_{\rm X}$) can be converted to the one defined by \citeauthor{Kellermann1989} using the formula
\begin{equation}
\label{eq_A2_RKconversion}
{\cal R}_{\rm K} = \frac{F_{\nu_5} / F_{\nu_{\rm r}}}{F_{\nu_{\rm B}}/F_{\nu_o}}
 \times \frac{(1+z)^{\alpha_{5} - \alpha_{\rm r}}}{(1+z)^{\alpha_{\rm B} -\alpha_{\rm o}}}
 \times \frac{L_{\nu_{\rm r}}}{L_{\nu_{\rm o}}}.
\end{equation}

Defining the two-point spectral indeces,
\begin{equation}
\label{eq_A3_alphar}
\langle \alpha_r \rangle = - \frac{\log \left( F_{\nu_5}/F_{\nu_r} \right)}{\log \left( \nu_5/\nu_r \right)}
\end{equation}
and
\begin{equation}
\label{eq_A4_alphao}
\langle \alpha_{o} \rangle = - \frac{\log \left( F_{\nu_{\rm B}}/F_{\nu_{o}} \right)}{\log \left( \nu_{\rm B}/\nu_{o} \right)},
\end{equation}
and noting that for $z<2$ and typical radio and optical/infrared spectral indeces the term $(1+z)^{(\alpha_{5} - \alpha_{\rm r})-(\alpha_{\rm B} -\alpha_{\rm o})}$ is of the order of unity, one can find that
\begin{equation}
\label{eq_A5_RKfinal}
{\cal R}_{\rm K} \simeq \frac
{(\nu_5/\nu_{\rm r})^{-\langle \alpha_{\rm r} \rangle}}
{(\nu_{\rm B}/\nu_{\rm o})^{-\langle \alpha_{\rm o} \rangle}} 
  \times \frac{L_{\nu_{\rm r}}} {L_{\nu_{\rm o}}}.
\end{equation}

Using this formula for the \citeauthor{Gurkan2019} and $0.5-0.7$ QSOs samples (Section~\ref{sec_2_RLFvsz} and \ref{sec_3_RLFvsMBHandEddratio}) with $\langle \alpha_r \rangle = 0.7$ \citep[see][]{Yuan2018} and $\langle \alpha_o \rangle = 0.5$ \citep[see][]{Richards2006} we get $\mathcal{R}_{\rm K}^{\rm (G)} \simeq 0.12 \times L_{\nu_{144}}/L_{\nu_i}$ and $\mathcal{R}_{\rm K}^{\rm (0.5-0.7)} \simeq 0.59 \times L_{\nu_{1.4}}/L_{\nu_i}$.

For the sample of \textit{Swift}/BAT AGNs taken from \citet{Rusinek2020} the two-point spectral index of $\langle \alpha_o \rangle$ between $\nu_{\rm W3}$ and $\nu_{\rm B}$ is on the order of $\sim 1$ \citep[see the SED in][]{Gupta2020} and then $\mathcal{R}_{\rm K}^{\rm (R)} \simeq 11 \times L_{\nu_{1.4}}/L_{\nu_{\rm W3}}$.

\section{$0.5-0.7$ QSOs sample}
\label{appendix_B_sample0507}

Table~\ref{tbl_appendix_whole_sample}.1 presents the most important information about some of the sources in the $0.5-0.7$ QSOs sample. The complete sample is available as supplementary material online. 

\renewcommand{\thetable}{\Alph{section}.1}
\begin{table*}[h]
\label{tbl_appendix_whole_sample}
\begin{center}
\caption{Sample of $0.5-0.7$ QSOs (total of 3511 objects) used in this study.}
\hspace{-2.5cm}
\begin{tabular}{lccccrrcc} \hline
SDSS Name & $z$ & $\log M_{\rm BH}$ & $\log L_{\rm bol}$ & $\log \lambda_{\rm Edd}$ & \mc{$F_{\rm 1.4}$} & $L_{\nu_{1.4}}/L_{\nu_i}$ & Radio Class$^a$ & Radio Morphology$^b$ \\ 
  &  & ($M_{\odot}$) & (erg\,s$^{-1}$) &  & \mc{(mJy)} &  &  &  \\ \hline \hline
081214.31+063653.2 & 0.6603 & 8.73 & 45.89 & -0.94 &   1.0 &    7.60 & RQ & U  \\
081218.40+110300.5 & 0.6432 & 9.27 & 45.58 & -1.79 &   2.1 &   31.33 & RI & C  \\
081259.17+150226.0 & 0.6048 & 8.76 & 45.50 & -1.36 &   1.0 &   15.16 & RQ & U  \\
081259.76+211103.9 & 0.6123 & 8.88 & 45.42 & -1.56 &   1.0 &   18.79 & RQ & U  \\
081318.85+501239.7 & 0.5714 & 9.14 & 45.66 & -1.58 & 487.3 & 4532.62 & RL & L  \\
081322.58+171638.6 & 0.6121 & 8.79 & 45.71 & -1.18 &   1.0 &    9.57 & RQ & U  \\
081327.61+561625.8 & 0.5063 & 8.77 & 45.61 & -1.26 &   4.2 &   32.64 & RI & C  \\
081344.01+171103.2 & 0.5814 & 9.30 & 45.54 & -1.86 &  27.8 &  350.94 & RL & E  \\
081416.99+252935.1 & 0.6026 & 9.57 & 45.50 & -2.17 &   1.0 &   14.93 & RQ & U  \\
081502.81+513313.3 & 0.5128 & 8.54 & 45.39 & -1.25 &   1.0 &   13.37 & RQ & U  \\
081510.31+403750.5 & 0.5895 & 8.53 & 45.44 & -1.19 &   1.0 &   16.21 & RQ & U  \\
081512.01+115311.6 & 0.5511 & 8.55 & 45.61 & -1.04 &   1.0 &    9.42 & RQ & U  \\
081558.06+501232.2 & 0.5996 & 9.01 & 46.09 & -1.02 &   1.0 &    3.84 & RQ & U  \\
081601.94+240925.8 & 0.5529 & 9.16 & 45.63 & -1.63 &   1.0 &    9.09 & RQ & U  \\
081615.35+550615.5 & 0.6145 & 8.62 & 45.52 & -1.20 &   1.0 &   14.95 & RQ & U  \\
081629.13+493249.9 & 0.6846 & 8.99 & 45.75 & -1.34 &   1.0 &   11.44 & RQ & U  \\
081632.66+152008.2 & 0.6586 & 8.95 & 45.44 & -1.61 &   1.0 &   21.15 & RQ & U  \\
081655.34+074311.5 & 0.6442 & 8.53 & 45.65 & -0.98 &   1.0 &   12.51 & RQ & U  \\
081707.04+351819.8 & 0.694  & 8.89 & 45.43 & -1.56 &   1.0 &   24.56 & RI & C  \\
081710.05+321802.1 & 0.6159 & 8.60 & 45.33 & -1.37 &   1.0 &   23.33 & RI & U  \\
 \hline  
\end{tabular}
\end{center}
\vspace{-0.2cm}
\textbf{Notes.} This subset of the table demonstrates format and content. \\
$^a$ RL -- radio-loud; RI -- radio-intermediate; RQ -- radio-quiet. \\
$^b$ L -- lobed; E -- extended (but not lobed); C -- compact; U -- undetected.  
\end{table*}

\newpage 
\newpage 

\bibliographystyle{aasjournal}
\bibliography{refs}

\end{document}